%\magnification \magstep1
%\baselineskip 22truept
%\nopagenumbers

\font\bbf=cmbx12

\font\srm=cmr9
\font\sit=cmti9
\font\sbf=cmbx9

\def \Ezp	{{\bf E}^{zp}}
\def \Bzp	{{\bf B}^{zp}}
\input epsf.tex
\
%\nopagenumbers
{\srm \hfill $\copyright$ 2003 The American Institute of Physics}
\bigskip
\centerline{\srm To be published in AIP Conference Proceedings of the}
\centerline{\srm Space Technology and Applications International Forum
(STAIF-2003)}\par
\centerline{\sit Expanding the Frontiers of Space}\par
\centerline{\srm February 2--6, 2003, Albuquerque, NM}\par
\bigskip

\centerline{\bbf Update on an Electromagnetic Basis for Inertia, Gravitation, }
\centerline{\bbf the Principle of Equivalence, Spin and Particle Mass Ratios}
\bigskip
\centerline{Bernard Haisch$^1$, Alfonso Rueda$^2$, L. J. Nickisch$^3$, and Jules
Mollere$^4$}
\bigskip
\centerline{\sit $^1$Calif. Inst. for Physics \& Astrophysics, 901 Mariners Island Blvd., Ste. 325, San Mateo, CA 94404}
\centerline{\sit $^2$Dept. of  Electrical Eng., California State Univ., Long Beach, CA
90840}
\centerline{\sit $^3$Mission Research Corp., Monterey, CA 93940-5776}
\centerline{\sit $^4$Henderson State Univ., Arkadelphia, AR 71999-0001}
\centerline{\sit $^1$haisch@calphysics.org, 650-593-8581, fax:
650-595-4466, $<$www.calphysics.org$>$}
\bigskip
\font\bbf=cmbx10

{\parindent 0.2truein \narrower \noindent
{\sbf Abstract.}
{\srm A possible connection between the electromagnetic quantum vacuum and inertia was 
first published by Haisch, Rueda and Puthoff (1994). If correct, this would
imply that mass may be an electromagnetic phenomenon and thus in principle subject to
modification, with possible technological implications for propulsion. A multiyear
NASA-funded study at the Lockheed Martin Advanced Technology Center further
developed this concept, resulting in an independent theoretical validation of the
fundamental approach (Rueda and Haisch, 1998ab). Distortion of the quantum
vacuum in accelerated reference frames results in a force that appears to account for
inertia. We have now shown that the same effect occurs in a region of curved spacetime,
thus elucidating the origin of the principle of equivalence (Rueda, Haisch and Tung, 2001).
A further connection with general relativity has been drawn by
Nickisch and Mollere (2002): zero-point
fluctuations give rise to spacetime micro-curvature effects yielding a complementary perspective on the
origin of inertia. Numerical simulations of this effect demonstrate the manner in which a massless
fundamental particle, e.g. an electron, acquires inertial properties; this also shows the apparent origin of
particle spin along lines originally proposed by Schr\"odinger. Finally, we suggest
that the heavier leptons (muon and tau) may be explainable as spatial-harmonic resonances of the (fundamental)
electron.  They would carry the same overall charge, but with the charge now having spatially lobed structure,
each lobe of which would respond to higher frequency components of the electromagnetic quantum vacuum, thereby
increasing the inertia and thus manifesting a heavier mass.}

}

\bigskip\centerline{\bbf BACKGROUND}
\medskip\noindent
It has been assumed since the late 1960s that an explanation for the origin of mass
compatible with the Standard Model of particle physics must involve a postulated Higgs
field.  The idea is that particles are presumed to acquire the property of mass through
interactions with an underlying universal field called the Higgs field, which is
carried by the Higgs bosons (as the electromagnetic field is carried by photons).
Evidence for the existence of a Higgs field has thus been sought in attempts to create
and detect Higgs bosons in colliders by concentration of a sufficiently
large amount of energy in particle collisions.

\medskip\noindent
As the Large Electron-Positron Collider (LEP) at
CERN approached its final shutdown (to make way for construction of the Large Hadron
Collider) in the autumn of 2000, the four LEP detector groups were reporting tentative
evidence for a Higgs boson with a mass near 115 GeV/$c^2$, at the very limit of the LEP
collider (which was already being pushed beyond its design limits). This was a tantalizing
possibility, but the enhancement above background of the measurements was less than 3
standard deviations. However further analysis of the LEP data which had been thought
to show evidence of the short-lived presence of a Higgs boson failed to demonstrate any
convincing signal. This led to the publication of an article in the December 5, 2001 issue
of the popular British magazine {\it New Scientist} with the headline ``No sign of
the Higgs boson'' together with a strong suggestion that the Higgs does not exist. This
anti-Higgs story was widely picked up in the press. However as
pointed out by the CERN groups in a dissenting letter also published in that
magazine, the mass of the Higgs boson could lie as high as 200 GeV/$c^2$.

\medskip\noindent
Even if the elusive Higgs field is found, this may not shed any light on the property of inertia. As our approach
shows, it is possible and potentially fruitful to question whether inertia is simply an innate property
of matter, or whether it can be shown to have a separate and specific origin whose mechanism may be altered
experimentally (see the discussion of such issues in the monograph by Jammer, 2000).
The Standard Model does not address the question of why
and how the structure of an elementary particle is able to display
inertial effects.  For a more extensive discussion of this and related points
we refer to Dobyns, Rueda and Haisch (2000) and Haisch, Rueda and Dobyns
(2001). The case of gravity is exactly parallel to the situation of
inertia as should be expected in light of the principle of equivalence.
The proposal that the quantum vacuum (also known in its more
restrictive  version as the electromagnetic zero-point field, or ZPF) sits
at the origin of inertia and gravity, is not only an interesting
theoretical proposition in itself but it is also technologically
relevant. It opens the possibility to perhaps manipulate inertial and gravitational forces. This would have
revolutionary implications for new technologies, especially for space
exploration.

\medskip\noindent
In this paper we  present a summary of work in progress which indicates how the electromagnetic
quantum vacuum interacting with idealized massless particles via the phenomenon of
{\it zitterbewegung} may explain or at least provide deeper insight into the following
fundamental laws and properties of matter:

-- origin of inertia (i.e. why does {\bf F} equal $m{\bf a}$ in Newton's equation of
motion?)

-- principle of equivalence for inertial and gravitational mass

-- de Broglie wavelength of material objects (in particular the electron)

-- nature of spin

-- mass ratios of fundamental particles (in particular the electron, muon and tau lepton)

\medskip\noindent
Earlier papers on these topics are online at $<$http://www.calphysics.org/sci\_articles.html$>$.

\bigskip\bigskip\centerline{\bbf THE ELECTROMAGNETIC QUANTUM VACUUM (ZPF)}

\medskip\noindent
The Heisenberg uncertainty relation applied to a
harmonic oscillator requires that its ground state have a non-zero energy of $h\nu/2$, since quantum
mechanically a particle cannot simultaneously be exactly at the bottom of its potential well and have
exactly zero momentum. 
The quantization of the
electromagnetic field in terms of quantum-mechanical operators is found in standard
textbooks. As stated by Loudon (1984): ``The electromagnetic field is now quantized by the
association of a quantum-mechanical harmonic oscillator with each mode {\bf k} of the radiation
field." Thus there exists
the same
$h\nu/2$ zero-point energy expression for each mode of the electromagnetic field as for a mechanical
oscillator. 
Summing up the energy over the modes for all frequencies, directions, and polarization
states, one arrives at a zero-point energy density for the electromagnetic fluctuations, and this is
the origin of the electromagnetic quantum vacuum. An energy of $h\nu/2$ per mode of the field
characterizes the fluctuations of the quantized radiation field in quantum field theory. In the
semi-classical representation of stochastic electrodynamics (SED, see the monographs by de la Pe\~na and Cetto,
1996 and Milonni, 1994) the quantum vacuum is represented by propagating electromagnetic plane waves, $\Ezp$ and
$\Bzp$, of random phase having this average energy,
$h\nu/2$, in each mode.

\medskip\noindent
The volumetric density of modes
between frequencies $\nu$ and
$\nu+d\nu$ is given by the density of states function
$N_{\nu}d\nu=(8\pi\nu^2/c^3)d\nu$. Each state has a minimum $h\nu/2$
of energy,
and using this density of states function and this minimum zero-point energy per state one gets the
spectral energy density of the electromagnetic quantum vacuum:
$$\rho(\nu)d\nu={8\pi\nu^2 \over c^3} {h\nu
\over 2} d\nu .
\eqno(1)
$$
Writing this zero-point radiation together with ordinary blackbody radiation, the energy density is:
$$\rho(\nu,T)d\nu={8\pi\nu^2 \over c^3} \left( {h\nu \over e^{h\nu /kT} -1}
+{h\nu
\over 2}\right) d\nu .
\eqno(2)
$$
The first term (outside the parentheses) represents the mode density, and the
terms inside the parentheses are the average energy per mode of thermal
radiation at temperature $T$ plus the zero-point energy, $h\nu/2$. Take away all thermal energy by
formally letting $T$ go to zero, and one is still left with the zero-point term. The laws of quantum
mechanics as applied to electromagnetic radiation force the existence of a
background sea of electromagnetic zero-point energy that is traditionally called the electromagnetic
quantum vacuum.

\medskip\noindent
It was discovered in
the mid-1970's that the quantum vacuum acquires special
characteristics when viewed from an accelerating frame.
Just as there is an event horizon for a black hole, there is an analogous event
horizon for an accelerating reference frame. Similar to radiation from evaporating black holes
proposed by Hawking (1974), Unruh (1976) and Davies (1975)
determined that a
Planck-like radiation component will arise out of the quantum vacuum in a uniformly-accelerating
coordinate system having constant proper
acceleration {\bf a} (where $|{\bf a}|=a$) with what amounts to an effective
``temperature''
$$
T_a = {\hbar a \over 2 \pi c k} .  \eqno(3)
$$
This ``temperature'' characterizing Unruh-Davies radiation does not originate in emission from
particles undergoing thermal
motions.
As discussed by Davies, Dray and Manogue (1996):

\medskip\noindent
{\parindent 0.4truein \narrower \noindent
One of the most curious properties to be discussed in recent years is the
prediction that
an observer who accelerates in the conventional quantum vacuum of Minkowski
space will
perceive a bath of radiation, while an inertial observer of course
perceives nothing. In
the case of linear acceleration, for which there exists an extensive
literature, the
response of a model particle detector mimics the effect of its being
immersed in a bath of
thermal radiation (the so-called Unruh effect).

}

\medskip\noindent
This ``heat bath'' is a quantum phenomenon. The ``temperature'' is
negligible for most
accelerations. Only in the extremely large gravitational fields of black
holes or in
high-energy particle collisions  can this  become significant. At the June 2000 meeting of the American
Astronomical Society, P. Chen of the Stanford Linear Accelerator Center proposed using an ultra high
intensity laser to accelerate electrons violently enough to directly detect Unruh-Davies radiation.

\medskip\noindent
Unruh and Davies treated the electromagnetic quantum vacuum as a scalar field. If a true vectorial
approach is considered there appear additional terms beyond the quasi-thermal Unruh-Davies component.
For the case of no true external thermal radiation
$(T=0)$ but including the acceleration effect
$(T_a)$, eqn.~(1) becomes (Boyer 1980)
$$\rho(\nu,T_a)d\nu = {8\pi\nu^2 \over c^3}
\left[ 1 + \left( {a \over 2 \pi c \nu} \right) ^2 \right]
\left[ {h\nu \over 2} + {h\nu \over e^{h\nu/kT_a}-1} \right] d\nu . \eqno(4)
$$
While these acceleration-dependent terms do not show any spatial
asymmetry in
the expression for the spectral energy density, an asymmetry does
appear when the
momentum flux of this radiation is calculated, resulting in a non-zero flux.
This  appears to be the process underlying inertial and gravitational forces.

\bigskip\bigskip\centerline{\bbf ZITTERBEWEGUNG}

\medskip\noindent
In his study of the coordinate operator in the Dirac
equation, Schr\"odinger (1930, 1931) discovered microscopic oscillatory
motion at the speed of light, which he called {\it zitterbewegung}.  While Dirac argued that such
motion does not violate relativity or quantum theory (see Dirac, 1958), from a classical particle point
of view, these speed of light motions would seem to imply
masslessness of the particle. Dirac theory also describes particle
spin, and Schr\"odinger considered spin to be an orbital angular
momentum that is a consequence of the vacuum fields.  This
view of spin was explored further by Huang (1952) and Barut and
Zanghi (1984).

\medskip\noindent
We take the view presented in the monograph by de la Pe\~na and Cetto (1996) that {\it zitterbewegung}
is the result of electromagnetic quantum vacuum fluctuations acting upon a fundamentally point-like
massless charged particle. We discuss below how such a particle constantly undergoing transverse
changes in direction due to these fluctuations would manifest the property of inertia, hence appearing to
possess the property of mass. {\it The spatially averaged size of a point-like particle undergoing such
perturbations corresponds to the Compton radius} (see MacGregor, 1992), thus showing one of several connections
between quantum properties and {\it zitterbewegung} (Hestenes, 1990). Spin and the de Broglie wavelength can
also be understood from this perspective.

\bigskip\bigskip\centerline{\bbf THE LORENTZ FORCE APPOACH TO INERTIA (HRP)}

\medskip\noindent
In the paper ``Inertia as a zero-point field Lorentz force'' Haisch, Rueda and Puthoff (HRP, 1994)
assumed that a fundamental particle (such as an electron) could be treated as a two-dimensional Planck oscillator
driven the by electric components ($\Ezp$) of the ZPF to oscillate in the $xy$-plane. They then examined the
effects of the magnetic components ($\Bzp$) of the ZPF on the Planck oscillator under the condition of constant
acceleration in the
$z$-direction. The result was that the Lorentz force due to $\Bzp$ fluctuations proved to be proportional to the
acceleration of the Planck oscillator, thus suggesting its interpretation as the reaction force due to inertia.

\bigskip\bigskip\centerline{\bbf THE POYNTING VECTOR APPROACH TO INERTIA (RH)}
\medskip\noindent
The approach by Rueda and Haisch (RH) relies on making standard transformations of the
$\Ezp$ and $\Bzp$ from a stationary to an accelerated coordinate
system. In a stationary or uniformly-moving frame
the
$\Ezp$ and
$\Bzp$ constitute an isotropic radiation pattern. In an accelerated frame the
radiation pattern acquires asymmetries. There turns out to be a non-zero Poynting vector in
any accelerated frame, and this carries a non-zero net flux of electromagnetic momentum. The
scattering of this momentum flux generates a reaction force, ${\bf F}_r$. RH found an invariant
scalar with the dimension of mass describing the resistance to acceleration resulting from this
process. We interpret this scalar as the inertial mass,
$$
m_i={V_0 \over c^2} \int \eta(\nu) \rho_{zp}(\nu) \ d\nu , \eqno(5)
$$
where $\rho_{zp}$ is the well known spectral energy density of the electromagnetic quantum vacuum  of
eqn.~(1). In other words, the amount of electromagnetic zero point energy instantaneously transiting
through an object of volume $V_0$ and interacting with the quarks and electrons in that object is
what constitutes the inertial mass of that object. It is change in the momentum of the radiation
field that creates the resistance to acceleration usually attributed to the inertia of an object.

\medskip\noindent
Indeed, not only does the ordinary form of Newton's second law, ${\bf F}=m_i{\bf a}$, emerge
from this analysis, but one can also obtain the relativistic form of the second law:
$${\cal F}={d{\cal P} \over d\tau} = {d \over d\tau} (\gamma_{\tau} m_i c, \ {\bf p}
\ ) .
\eqno(6)
$$
The origin of inertia, in this picture, becomes remarkably intuitive. Any material object
resists acceleration because the acceleration produces a perceived flux of radiation
in the opposite direction that scatters within the object and thereby pushes against
the accelerating agent. Inertia in the present model appears as a kind of acceleration-dependent
electromagnetic quantum vacuum drag force acting upon electromagnetically-interacting elementary
particles (electrons and quarks).
The relativistic law for ``mass" transformation --- that is, the formula
describing how the {\it inertia} of a body has been calculated to change
according to an observer's relative motion --- is automatically satisfied
in this view, because the correct relativistic form of the reaction force
is derived, as shown in eqn.~(6).

\bigskip\bigskip\centerline{\bbf THE NEW CONNECTIVITY APPROACH}

\medskip\noindent
Both of the approaches above (together called the RHP approach for convenience) assume classical electrodynamics
operating in flat spacetime.  Einstein's field equations for general relativity (GR)

$$G_{\mu \nu} = 8 \pi T_{\mu \nu} \eqno{(7)}$$

\noindent
describe how curved spacetime geometry ($G_{\mu \nu}$) is produced by the presence of matter or energy
as described by the energy-momentum tensor ($T_{\mu \nu}$).  Nickisch and Mollere have considered the possibility
that electromagnetic fields, including that of the zero-point fluctuations, can be treated as a distortion in
the spacetime of the charge.  A massless charge would behave like a photon, following a null geodesic, but in a
spacetime defined by electromagnetic fields.  A photon, in the absence of any energy or matter other than the
zero-point fluctuations, will follow an unperturbed flat spacetime trajectory.  However unlike the spacetime of
a photon, the spacetime of the massless charge is defined by the distortions of the zero-point fluctuations,
producing a geodesic description of {\it zitterbewegung}.  Additional electromagnetic fields may produce a
non-zero-mean drift of the {\it zitterbewegung} that is also accounted for in the geodesic motion.  These
non-zero-mean effects ``accumulate'' into a stretching of the particle's spacetime, and this stretching is
perceived by external observers to be inertia.

\medskip\noindent 
In the Nickisch-Mollere ``Connectivity'' theory, the ZPF defines a curvature in the particle's spacetime. The
metric describing this curvature implies a transformation to the
viewpoint of an observer who assumes spacetime is flat (the
Connective transformation). Application of the Connective
transformation produces the usual effects of inertia when observed
in Minkowski (flat) spacetime, including hyperbolic motion in a
static electric field (above the vacuum) and uniform motion
following an impulse. The motion of the massless charge is a helical
motion that can be equated to the particle spin of quantum theory.
This spin has the properties expected from quantum theory, being
undetermined until ``measured'' by applying a field, and then
being found in either a spin up or spin down state.

\medskip\noindent
In Connectivity it is assumed that the equation of motion of a
massless charge is that of a geodesic in a spacetime whose
curvature is defined by the electromagnetic fields encountered by
the particle. Since the charge is assumed to be massless and
moving at the speed of light (following a ``null'' curve, in the
terminology of relativity), proper time cannot be used as the
affine parameter of the geodesic (proper time intervals vanish for
null geodesics). However, normal time serves well as an affine
parameter. The equation of the geodesic is therefore taken to be

$${dp^{\mu} \over dt} + {1 \over m_*} \Gamma^{\mu}_{\nu \rho} p^{\nu} p^{\rho} = 0 \ ,
\eqno{(8)}
$$
where
$$p^{\mu}=m_* c n^{\mu} \ , \eqno{(9)}
$$
and
$$
n^{\mu}=(n^0,{\bf n}) \ . \eqno{(10)}
$$
Here $p^{\mu}$ is the four-momentum of the charge $q$, $c$ is
the speed of light, $F^{\mu \nu}$ is the electromagnetic field
tensor of the impressed fields including the ZPF, and $n^{\mu}$ is
the direction vector of the particle motion. The mass parameter
$m_*$ has the dimensions of mass, but it is {\it not} mass;
particles described by eqn.~(8) move at the speed of light.
$\Gamma^{\mu}_{\nu \rho}$ are the Christoffel symbols of the second
kind. The connection terms (the terms containing the Christoffel
symbols) are equated with the Lorentz force, thus showing the connection between this approach and that of HRP.
That is,
$$
\Gamma^{\mu}_{\nu \rho} p^{\nu} p^{\rho} = -{q \over c} F^{\mu}_{\nu} p^{\nu} n^0 \ .
\eqno{(11)}
$$
These equations can be solved for the metric $g_{\mu \nu}$ of
the particle's spacetime, though not uniquely.  The equations
(11) actually define a class of metrics.  Further
constraints are required to select a particular solution from this
class.  In particular, the geodesic in the particle spacetime
should be a null curve as expected for a massless object. One interesting aspect of the metrics derived from
eqn.~(11) is that they turn out to depend on the particle history through time integrations.  Since each
particle experiences its own history of encountered fields (including zero-point fluctuations), this means that
the local spacetime distortion observed by a particle is unique to it; two particles at the same place and time
will, in general, see spacetime stretched in a somewhat different way.  This, in Connectivity, is how two
identical massless charges can have different momenta.

\medskip\noindent
Using eqn.~(11), the geodesic equation (8)  can be
written as,
$$
{dn^j \over dt} = {q \over m_* c} \left[ F_{\nu}^j n^{\nu} n^0 - F_{\nu}^0 n^{\nu}
n^j \right ] + {n^j \over n^0} {dn^0 \over dt} \ , \eqno{(12)}
$$
$$
{dm_* \over dt} = {q \over c} F_{\nu}^0 n^{\nu} - {m_* \over n^0} {dn^0 \over dt} \ .
\eqno{(13)}
$$
In general $n^0$ does not retain a value of unity, but
changes in a way that preserves the null curve property,
$$
g_{\mu \nu} p^{\mu} p^{\nu} = 0 \ . \eqno{(14)}
$$
Note that eqn.~(13), which is the zeroth equation of
eqn.~(8), is an equation for the parameter $m_*$.  Thus $m_*$
is not a constant, but rather varies in response to applied
forces. The effect is to introduce time dilation (or Doppler
shifting) in the energy-momentum four vector analogous to the
gravitational redshift of GR.

\medskip\noindent
Eqns.~(8--14) describe the motion of a massless
charge in response to impressed electromagnetic fields. The charge
moves at a constant speed (the speed of light) with a changing
direction given by eqn.~(12). When the impressed fields include
the ZPF, this motion may be regarded as Schr\"odinger's
{\it zitterbewegung}. When a field above the vacuum is applied, the
charge will be observed to drift in a preferred direction in its
{\it zitterbewegung} wander. This is illustrated in Figure 1, which
shows the trajectory of a massless charge computed from eqn.~(8) using one of the metrics from the class of
metrics implied by eqn.~(11). The electromagnetic fields influencing
the motion of the charge are a random realization of the ZPF with
a superimposed uniform electric field in the vertical direction
(the driving field). Note that the charge drifts upward in
response to the driving field. We see that the ZPF drives the
charge in a pseudo-helical motion as in Schr\"odinger's orbital
angular momentum explanation for spin.

\medskip\noindent
The metric $g_{\mu\nu}$ implies a transformation to Minkowski
(flat) spacetime, the so-called Connective transformation. The
transformation $C_\mu^\nu$ from the particle's spacetime to
Minkowski spacetime is related to the metric $g_{\mu\nu}$ by
$g_{\mu\nu}=C^\rho_\mu C^\sigma_\nu \eta_{\rho\sigma}$, or
$$
g=C\cdot\eta\cdot\tilde{C} \ , \eqno{(15)}
$$
where $\tilde{C}$ is the transpose of $C$ and $\eta$ is the flat spacetime metric.

\topinsert
\centerline{\epsfxsize=1.9in \epsfbox{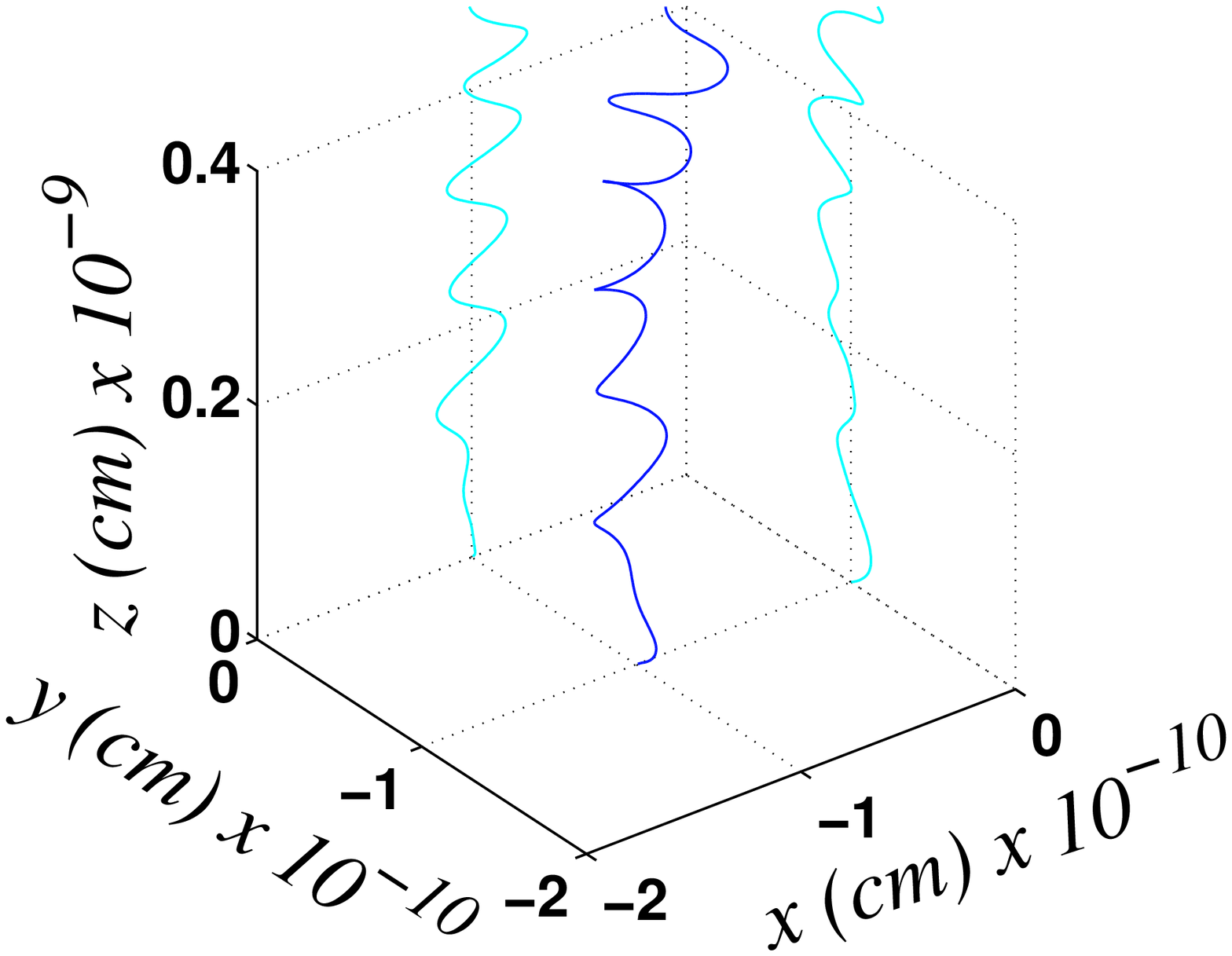} \epsfxsize=1.9in \epsfbox{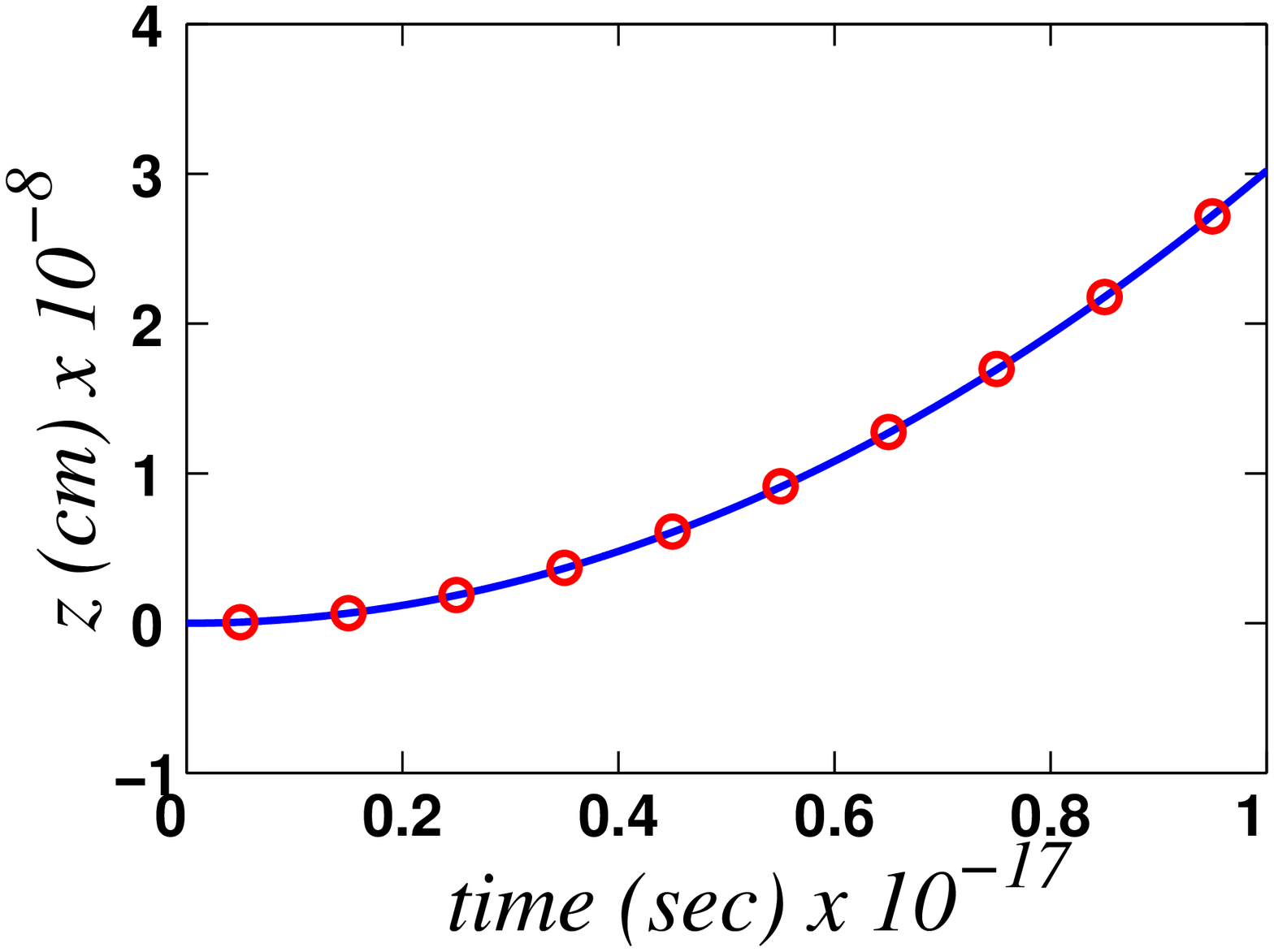} \ \ \ \ \epsfxsize=1.9in
\epsfbox{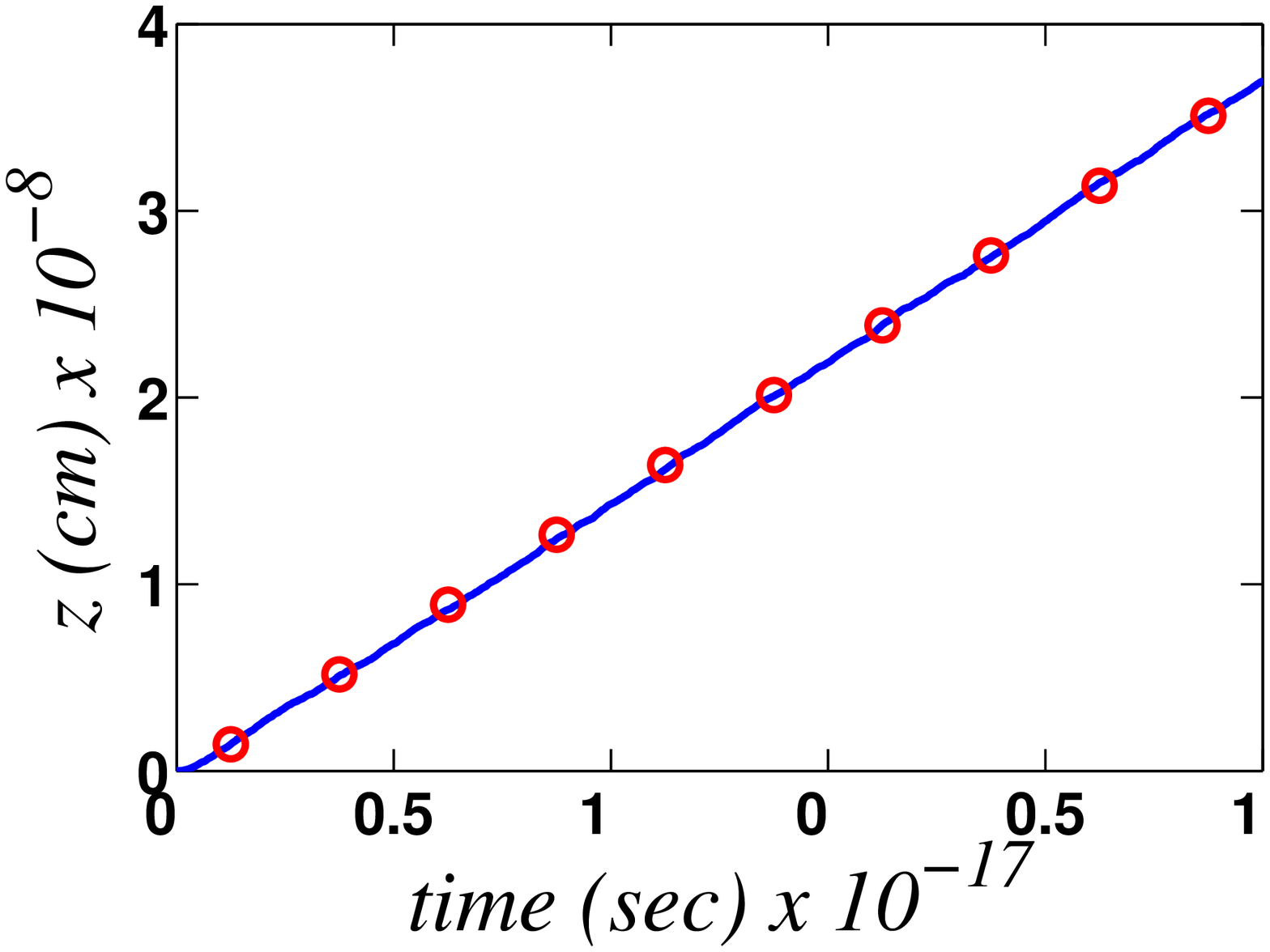} \ \ \ \ }
\medskip
\hbox{\srm \raise 0.16 truein \vbox{\hsize=2.0truein \noindent {\sbf FIGURE 1.} Massless Charge Trajectory in a
Uniform Electric Field plus the ZPF, Showing Spin-like Orbital \ Motion \ and \ its \ Projections 

\noindent onto the $xz$ and $yz$
Planes.}

\kern 0.065truein \vbox{\hsize=2.1truein \noindent {\sbf FIGURE 2.} Massless Charge Motion in a Uniform Electric
Field plus the ZPF, Obtained Using Connectivity (solid curve), compared to 
the Hyperbolic Motion
of a Massive Charge in Special Relativity (circles).}

\kern 0.065truein \raise 0.16 truein \vbox{\hsize=2.2truein \noindent {\sbf FIGURE 3.} Massless Charge Motion in the ZPF
Following an Impulse, Obtained Using Connectivity (solid curve), Compared to Constant Speed Motion (circles).}}
\endinsert

\medskip\noindent
It is the application of the Connective transformation that allows
one to view the particle trajectory in Minkowski spacetime, and
here the effects of inertia appear. Two of the simplest
manifestations of inertia are hyperbolic motion of a charge in the
presence of a uniform electric field and uniform motion following
an impulse. Figure 2 displays the component of the motion of a
charge in the direction of a uniform electric field applied above
the ZPF, obtained by solving eqn.~(8) and applying the
Connective transformation to view the result in Minkowski
spacetime.  The solid curve is the result of the Connectivity
simulation.  The circles lie on the hyperbola defined by a massive
charge undergoing uniform acceleration in special relativity.  The
agreement is striking.  Viewed in Minkowski spacetime, the
massless charge is seen to accelerate hyperbolically as though it
had inertia.  Figure 3 is a similar depiction for the case in
which an impulse has been applied to the charge, and the charge is
observed in Minkowski spacetime to continue in uniform motion
following the impulse.  Here the circles lie on a straight line,
indicating that the particle travels at a constant speed following
the impulse.  Note that it is the average motion of the charge
that moves uniformly. Deviations about the average motion are
apparent and are, in fact, {\it zitterbewegung} driven by the ZPF.  The
average motion defines a timelike curve, as expected of a massive
particle.  This dynamic is well-modelled by a moving center of
mass with speed-of-light motions of the charge center about it,
and is therefore consistent with the successful ``displaced charge
center'' model that is often invoked to describe the
{\it zitterbewegung} of classical charges (see, e.g., Rueda, 1993).  The
beauty of Connectivity is that the full dynamics of the displaced
charge center model is obtained without having to assume particle
structure, or rather, the particle structure of that model is
obtained in a natural way.

\medskip\noindent
When the
forces acting on the charge are the Lorentz forces due to the
electromagnetic vacuum fields (the ZPF), these drive the charge in
{\it zitterbewegung} motion at the speed of light, in agreement with the
speed-of-light eigenvalues of the Dirac theory. When the charge
moves with a large average velocity in some direction, the
{\it zitterbewegung} motion extends to a quasi-helical motion that may
be the basis of particle spin.  This spin is undetermined until
``measured'' by applying a field that aligns the {\it zitterbewegung}
into helical motion, which will either be oriented with positive
or negative helicity (spin up or spin down, see Figure 1).

\medskip\noindent
The RHP theory of inertia has the
implication that potentially {\it all} mass is due to interaction
of bare massless charges with the vacuum fields, where ``charge''
is understood in the generalized sense as the charge associated
with any fundamental vacuum field.  
It has recently been argued that since the effects of the ZPF on a massless particle moving at the speed
of light can only involve transverse forces, no work can be done, hence no energy transferred from the
ZPF to the particle, thus contradicting the proposed RHP inertia-generating mechanism (Ibison 2001).
This problem is resolved in the Connectivity approach to inertia. Electromagnetic fields are assumed to describe
the curvature of spacetime, and massless charges simply follow
geodesic motion in this curved spacetime.  When the view is
transformed to the flat  spacetime assumption, inertial
forces appear.

\medskip\noindent
Since the interpretation of Connectivity is that electromagnetic
fields define a curvature of spacetime, it may be possible to show
that radiation is related to the distortion of spacetime required
to connect the views of different frames. This should then have
implications for the problem of the stability of atoms.
Furthermore, these distortions of spacetime are presumably no
different from that described by the usual gravitational form of
GR; it is simply a matter of scale.  The ZPF
represents spacetime distortions on a very fine scale.  However,
if there are many charges in a localized place (say in a star),
then their presence (as boundary conditions for the ZPF and
including their collective ZPF-induced radiation) changes the
normally isotropic ZPF to an EM-field distribution yielding a
broad overall spacetime curvature in that region. This broad
curvature may in fact be the spacetime curvature usually assigned
to gravity in GR, thereby providing the potential
to relate gravitational and inertial mass.

\topinsert
\epsfxsize=6.5in \epsfbox{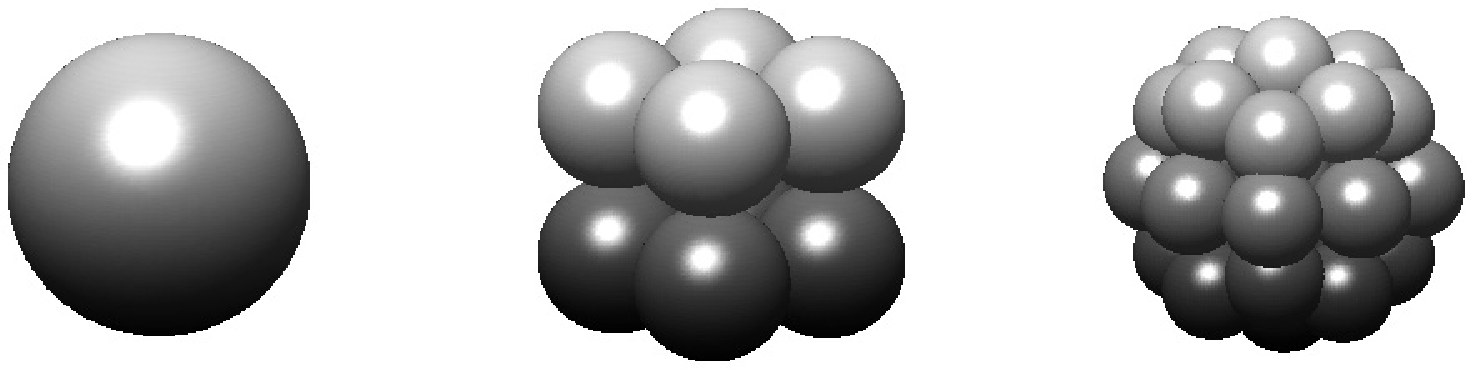} 
\noindent
\srm {\sbf FIGURE 4.} Notional Depiction of Charge Sub-volumes that Yield Lepton Mass Spectrum as Harmonic
Structures.
\endinsert

\medskip\noindent
We conclude with a conjecture that, while speculative, is at least
worthy of further study.  The RHP theory implies that the
effective inertial mass of a charge scales with either the third power of a resonance frequency or the fourth
power of a cutoff frequency for the charge's interaction with the ZPF.
If this cutoff is associated with a finite charge
distribution size, then it is reasonable to expect that the charge
distribution itself may be capable of exhibiting a harmonic mode
structure.  Now imagine that for some reason the first strongly
observable harmonic structure has the charge distribution
exhibiting a two-lobe structure in each of the three spatial
dimensions, with the second strongly observable harmonic
exhibiting a three-lobe structure in each of the three dimensions.
This is shown notionally in Figure 4. The number of charge
sub-volumes so generated scales as the cube of the cutoff
frequency of the fundamental. All told, then, the masses of the
higher states will scale as the sixth or the seventh power of a resonance or a cutoff
frequency respectively of the fundamental since each sub-lobe will be
responsive to proportionately higher frequencies of the ZPF (i.e.,
a sub-lobe with half the size of the fundamental charge
distribution size can be translated by wavelengths twice as small
as the fundamental charge distribution size).  When this
conjecture is applied to the lepton mass spectrum, we find for the
masses of the muon and tauon relative to the electron that $(m_\mu
/ m_e)^{1/7}=2.1$, $(m_\tau / m_e)^{1/7}=3.2$ for the case of a cutoff frequency or $(m_\mu
/ m_e)^{1/6}=2.4$, $(m_\tau / m_e)^{1/6}=3.9$ for the case of a resonance.  Form factors
associated with the actual shapes of the harmonic charge
structures could easily account for the small deviations of these
results from integer values.  The spin of each of these states is
associated with the {\it zitterbewegung} transport of the entire body as
a whole, so the spin of each of the leptons will be the same.
Although this mass relation is only conjecture, it provides an
indication of the potential for a vacuum field origin of inertia
to explain the mass spectra of elementary particles.

\bigskip\bigskip\centerline{\bbf ORIGIN OF WEIGHT AND THE WEAK EQUIVALENCE PRINCIPLE}

\medskip\noindent
Einstein introduced the {\it local Lorentz invariance} (LLI) principle in order to pass from special
relativity to GR. It is possible to use this principle immediately to extend the results of
the quantum vacuum inertia hypothesis to gravitation (details discussed in Rueda, Haisch and Tung, 2001).

\medskip\noindent
The idea behind the LLI principle is embodied in the Einstein elevator thought experiment. He
proposed that a freely-falling elevator in a gravitational field is equivalent to one that is not
accelerating and is far from any gravitating body. Physics experiments would yield the same
results in either elevator, and therefore a freely-falling coordinate frame in a gravitational field
is the same as an inertial Lorentz frame. (This is rigorously only true for a ``small
elevator'' since a gravitational field around a planet, say, must be radial, hence there are
inevitably tidal forces which would not be the case for an ideal acceleration.) The device Einstein
used to develop GR was to invoke an infinite set of such freely falling frames. In
each such frame, the laws of physics are those of special relativity. The additional features of
GR emerge by comparing the properties of measurements made in freely-falling Lorentz
frames ``dropped'' one after the other.

\medskip\noindent
This approach of Einstein is both elegant and powerful. The LLI principle immediately tells us that an
object accelerating through the electromagnetic quantum vacuum is equivalent to an object held fixed
in a gravitational field while the electromagnetic quantum vacuum is effectively accelerating
(falling) past it. The prediction of GR that light rays deviate from straight-line propagation in the
presence of a gravitating body
translates into acceleration (falling) of the electromagnetic quantum vacuum. An object accelerating
through the electromagnetic quantum vacuum experiences a flux which causes the inertia
reaction force. A fixed object past which the electromagnetic quantum vacuum is accelerating,
following the laws of GR, experiences the same flux and the resulting force is what we call
weight. That is why
$m_g=m_i$ and is the basis of the weak equivalence principle.

\bigskip\bigskip\centerline{\bbf CONCLUSIONS}

\medskip\noindent
It appears that a simple model of a particle as an electromagnrtic quantum vacuum-driven
oscillating charge with a resonance at its Compton frequency may simultaneously offer insight into
the nature of inertial and gravitational mass, the origin of the de Broglie wavelength and spin, and possibly particle mass
ratios.  For a recent popular-level discussion of our appoach to the nature of mass see the article by Chown (2001) in
{\it New Scientist}. Numerous papers are also online at $<$http://www.calphysics.org/sci\_articles.html$>$.

\bigskip\bigskip\centerline{\bbf APPENDIX: INERTIA AND THE DE BROGLIE WAVELENGTH}

\medskip\noindent
Four-momentum is defined as
%(Rindler, eqn.~27.4)
$$
{\bf P}= \left( {E \over c}, \ \ {\bf p} \right) = \left( \gamma m_0 c, \ \ {\bf p} \right)
=\left( \gamma m_0 c, \ \ \gamma m_0{\bf v} \right), \eqno({\rm A}1)$$
where $|{\bf P}|=m_0 c$ and $E=\gamma m_0 c^2$. The Einstein-de Broglie relation defines the
Compton frequency
$h \nu_C = m_o c^2$ for an object of rest mass $m_0$, and if we make the de Broglie
assumption that the momentum-wave number relation for light also characterizes matter
then ${\bf p}=\hbar {\bf k}_B$ where ${\bf
k}_B=2\pi(\lambda^{-1}_{B,1},\lambda^{-1}_{B,2},\lambda^{-1}_{B,3})$. We thus write
$$
{{\bf P} \over \hbar} = \left( {2\pi \gamma \nu_C \over c}, {\bf k}_B \right)
= 2 \pi \left( {\gamma \over \lambda_C}, {1 \over \lambda_{B,1}}, {1 \over \lambda_{B,2}}, {1
\over
\lambda_{B,3}} \right)
\eqno({\rm A}2)
$$
and from this obtain the relationship
$$
\lambda_B={c \over \gamma v} \lambda_C \eqno({\rm A}3)
$$
between the Compton wavelength, $\lambda_C$, and the de Broglie wavelength, $\lambda_B$. For a
stationary object $\lambda_B$ is infinite, and the de Broglie wavelength decreases in inverse
proportion to the momentum.

\medskip\noindent
Eqn.~(5) is very suggestive that quantum vacuum-elementary particle interaction involves
a resonance at the Compton frequency. 
de Broglie proposed that an elementary particle is associated with a localized
wave whose frequency is the Compton frequency.
As summarized by Hunter (1996): ``\dots what we regard as the (inertial) mass of the particle
is, according to de Broglie's proposal, simply the vibrational energy (divided by $c^2$)
of a localized oscillating field (most likely the electromagnetic field). From this
standpoint inertial mass is not an elementary property of a particle, but rather a
property derived from the localized oscillation of the (electromagnetic) field. de Broglie
described this equivalence between mass and the energy of oscillational motion\dots as
{\it `une grande loi de la Nature'} (a great law of nature).'' 

\medskip\noindent
This perspective is
consistent with the proposition that inertial mass, $m_i$, may be a coupling parameter between
electromagnetically interacting particles and the quantum vacuum. Although de Broglie assumed
that his wave at the Compton frequency originates in the particle itself (due to some
intrinsic oscillation or circulation of charge perhaps) there is an alternative interpretation
discussed in some detail by de la Pe\~na and Cetto that a particle  ``is tuned to a wave
originating in the high-frequency modes of the zero-point background field.'' The de Broglie
oscillation would thus be due to a resonant interaction with the quantum vacuum, presumably the same
resonance that is responsible for creating a contribution to inertial mass as in eqn. (5). In other
words, the electromagnetic quantum vacuum would be driving this
$\nu_C$ oscillation.

\medskip\noindent
We therefore suggest that an elementary charge driven to oscillate at the Compton
frequency, $\nu_C$,  by the quantum vacuum may be the physical basis of the $\eta(\nu)$
scattering parameter in eqn.~(5).  For the case of the electron, this would imply that
$\eta(\nu)$ is a sharply-peaked resonance at the frequency, expressed in terms of
energy, $h\nu_C=512$ keV. The inertial mass of the electron would physically be the reaction
force due to resonance scattering of the electromagnetic quantum vacuum radiation
at that frequency.

\medskip\noindent
{\it This leads to a surprising corollary.} It has been shown that as viewed from a
laboratory frame, a standing wave at the Compton frequency in the electron frame transforms
into a traveling wave having the de Broglie wavelength
for a moving electron. (Hunter, 1996; de la Pe\~na and Cetto, 1996; Kracklauer, 1992; Haisch and Rueda,
2000) The wave nature of the moving electron (as measured in the Davisson-Germer experiment, for example)
would be basically due to Doppler shifts associated with its Einstein-de Broglie resonance at the Compton
frequency. A simplified heuristic model shows this, and a detailed treatment showing the same result
may be found in de la Pe\~na and Cetto. Represent a quantum vacuum-like driving force field as two
waves having the Compton frequency $\omega_C=2\pi \nu_C$ travelling in equal and opposite
directions,
$\pm
\hat{x}$. The amplitude of the combined oppositely-moving waves acting upon an electron will be
$$
\phi=\phi_+ + \phi_{-}=2 \cos \omega_C t \cos k_C x. \eqno({\rm A}4)
$$
But now assume an electron is moving with velocity $v$ in the $+x$-direction. The wave
responsible for driving the resonant oscillation impinging on the electron from the front
will be the wave seen in the laboratory frame to have frequency $\omega_-=\gamma
\omega_C (1 - v/c)$, i.e. it is the wave below the Compton frequency in the laboratory
that for the electron is Doppler shifted up to the
$\omega_C$ resonance. Similarly the zero-point wave responsible for driving the electron resonant
oscillation impinging on the electron from the rear will have a laboratory frequency
$\omega_+=\gamma \omega_C (1 + v/c)$ which is Doppler shifted down to $\omega_C$ for the
electron. The same transformations apply to the wave numbers,
$k_+$ and $k_-$. The Lorentz invariance of the electromagnetic quantum vacuum spectrum ensures that
regardless of the electron's (unaccelerated) motion the up- and down-shifting of the laboratory-frame
spectral energy density will always yield a standing wave in the electron's frame.

\medskip\noindent
It can be shown that the superposition of these two oppositely-moving, Doppler-shifted waves is
$$
\phi'=\phi'_++\phi'_{-}=2 \cos(\gamma \omega_C t - k_B x) \cos(\omega_B t - \gamma k_C x).
\eqno({\rm A}5)
$$
Observe that for fixed $x$, the rapidly
oscillating ``carrier'' of frequency $\gamma \omega_C$ is modulated by the slowly varying
envelope function in frequency $\omega_B$. And {\it vice versa} observe that at a given $t$ the
``carrier'' in space appears to have a relatively large wave number $\gamma k_C$ which is
modulated by the envelope of much smaller wave number $k_B$. Hence
both timewise at a fixed point in space and spacewise at a given time, there appears a
carrier that is modulated by a much broader wave of dimension corresponding to the de
Broglie time $t_B=2\pi/\omega_B$, or equivalently, the de Broglie wavelength
$\lambda_B=2\pi/k_B$.

\medskip\noindent
de la Pe\~na and Cetto (1996) generalize this to include quantum vacuum radiation from all other
directions and conclude: ``The foregoing discussion assigns a physical
meaning to de Broglie's wave: it is the {\it modulation} of the wave formed by the Lorentz-transformed,
Doppler-shifted superposition of the whole set of random stationary electromagnetic waves of frequency
$\omega_C$ with which the electron interacts selectively.''
Another way of looking at the spatial modulation is in terms of the
wave function: the  spatial modulation of eqn.~(A5) is exactly the $e^{i p x / \hbar}$ wave
function of a freely moving particle satisfying the Schr\"odinger equation (Hunter, 1996).
In such a view the quantum wave function
of a moving free particle becomes a ``beat frequency'' produced by the relative motion of
the observer with respect to the particle and its oscillating charge.

\bigskip\bigskip\centerline{\bbf REFERENCES}

\medskip\noindent

{

\srm

\parskip=0pt plus 2pt minus 1pt\leftskip=0.25in\parindent=-.25in

Barut A. O. and Zanghi, N., {\sit Phys.~Rev.~Lett.} {\sbf 52}, 2009
(1984).
 
Boyer, T. H., {\sit Phys. Rev. D} {\sbf 21}, 2137 (1980).

Chown, M., {\sit New Scientist}, No. 2276, 3 Feb. (2001).

Davies, P. C. W., {\sit J. Phys. A} {\sbf 8}, 609 (1975).

de la Pe\~na, L. and Cetto, A.M., {\sit The Quantum Dice: An Introduction to Stochastic
Electrodynamics}, (Kluwer Acad. Publ.), (1996).

Davies, P. C. W., Dray, T. and Manogue, C. A. , {\sit Phys. Rev. D} {\sbf 53}, 4382 (1996).

Dirac, P. A .M.{\sit The Principles of Quantum Mechanics},
(Clarendon, Oxford), 4th edition, p.~262 (1958).

Dobyns, Y., Rueda, A. and B. Haisch, {\sit Found. Phys.}, {\sbf 30},  No.1, 59,
(2000)                                                   

Haisch, B., Rueda, A. and Dobyns, Y., {\sit Ann. Phys.}                       
(Leipzig), {\sbf 10}, No. 5, 393-414, (2001).                          

Haisch, B. and Rueda, A. {\sit Phys. Lett. A} {\sbf 268}, 224 (2000).

Haisch, B., Rueda, A. and Puthoff, H.E. (HRP), {\sit Phys. Rev. A} {\sbf 49}, 678 (1994).

Hawking, S. {\sit Nature} {\sbf 248}, 30 (1974).

Hestenes, D., {\sit Found. Phys.}, {\sbf 20}, No. 10, 1213 (1990).

Huang, K. {\sit Am.~J.~Physics} {\sbf 20}, 479 (1952).

Hunter, G. in {\sit The Present Status of the Quantum Theory of Light}, S. Jeffers et al. (eds.),
(Kluwer Acad. Publ.), chap. 12 (1996)

Ibison, M., www.arxiv.org/abs/physics/0106080 (2001).

Jammer, M. {\sit Concepts of Mass in Contemporary Physics and Philosopy}, Princeton Univ. Press
(2000).

Kracklauer, A. F., {\sit Physics Essays} {\sbf 5}, 226 (1992).

Loudon, R., {\sit The Quantum Theory of Light}, chap. 1,
(Oxford: Clarendon Press) (1983).

MacGregor, M., {\sit The Enigmatic Electron}, Kluwer, (1992).

Milonni, P.W.  {\sit The Quantum Vacuum}, Academic Press (1994).

Nickisch, L.~J. and J.~Mollere,
www.arxiv.org/abs/physics/0205086 (2002).

Rueda, A. {\sit Found.~Phys.~Lett.}, {\sbf 6}, no. 1, 75 and no. 2, 139
(1993).

Rueda, A. and Haisch, B., 
{\sit Physics Lett. A}, {\sbf 240}, 115 (1998a).
www.arxiv.org/abs/physics/9802031

Rueda, A. and Haisch, B., {\sit Found. Phys.}, {\sbf 28}, 1057 (1998b).
www.arxiv.org/abs/physics/9802030

Rueda, A., Haisch B. and Tung, R., www.arxiv.org/gr-qc/0108026, (2001).

Schr\"{o}dinger, E. {\sit Sitzungsb.~Preuss.
Akad.~Wiss.~Phys.-Math.~K1.} {\sbf 24}, 418 (1930); {\sbf 3}, 1
(1931).

Unruh, W. G., {\sit Phys. Rev. D} {\sbf 14}, 870 (1976).

}
\bye